\documentclass[12pt]{article}
\usepackage{graphicx,amsmath,amssymb}
                                                    
\newlength{\dinwidth}                                                    
\newlength{\dinmargin}                                                    
\def\lapproxeq{\lower .7ex\hbox{$\;\stackrel{\textstyle <}{\sim}\;$}}
\def\gapproxeq{\lower .7ex\hbox{$\;\stackrel{\textstyle >}{\sim}\;$}}
\def\be{\begin{equation}}                                                    
\def\ee{\end{equation}}                                                    
\def\bea{\begin{eqnarray}}                                                    
\def\eea{\end{eqnarray}}                                                    
\def\GeV{\rm GeV}

\begin{document}                                                    
\titlepage                                                    
\begin{flushright}                                                    
IPPP/08/94 \\
DCPT/08/188 \\
LTH 816 \\                                                    
29 May 2009 \\
\end{flushright}                                                    
                                                    
\vspace*{2cm}                                                    
                                                    
\begin{center}                                                    

{\Large \bf Generalised parton distributions at small $x$}\\

\vspace*{1cm} 

{\sc A.D. Martin}$^a$, {\sc C. Nockles}$^b$, {\sc
  M.G. Ryskin}$^{a, c}$, {\sc A.G. Shuvaev}$^c$ and {\sc T. Teubner}$^{b}$ \\

\vspace*{0.5cm}

$^a$ {\em Department of Physics and Institute for Particle Physics
  Phenomenology,\\ 

University of Durham, Durham DH1 3LE, U.K.}\\

$^b$ {\em Department of Mathematical Sciences,\\

 University of Liverpool, Liverpool L69 3BX, U.K.}

$^c$ {\em Petersburg Nuclear Physics Institute, Gatchina,
St.~Petersburg, 188300, Russia} \\

\end{center}

\vspace{8mm}

\begin{abstract}                                        
We justify the practical use of the Shuvaev integral transform approach to
calculate the skewed distributions, needed to describe diffractive
processes, directly from the conventional diagonal global parton
distributions. We address doubts which have been raised about
this procedure. We emphasise that the approach, on the one hand, satisfies all theoretical reqirements, and, on the other hand, is consistent with DVCS data at NLO.  We construct an easily accessible package for
the computation of these skewed distributions. 
\end{abstract}

\vspace{8mm}

\section{Motivation}
Skewed parton distributions are needed to
calculate exclusive diffractive production, such as $\gamma+p\to
VM + p$ for light or heavy vector mesons, see
e.g. \cite{MRT3,MRT2,Diehl:2007hd,Goloskokov:2007nt}, central
exclusive diffractive Higgs boson production at the LHC~\cite{KMR},
etc. For all these diffractive processes we need skewed parton distributions at
small values of $\xi \ll 1$, and small to medium scales. Data for diffractive
$J/\psi$ production, for example, test the gluon in the range $\xi \simeq x =
10^{-4} \ldots 5 \cdot 10^{-3}$ and effective scales $\mu^2 \sim 2
\ldots 8$ GeV$^2$~\cite{MNRT}. For exclusive Higgs production at the
LHC, the regime of $\xi \simeq x \simeq 10^{-2} $ and scales of the
order 5 GeV$^2$ will be relevant, see~\cite{KMR}.

\begin{figure}
\begin{center}
\includegraphics[width=0.5\textwidth]{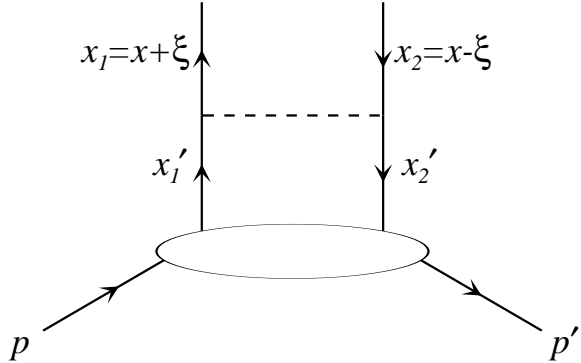}
\vspace{-3cm}
\caption{A schematic diagram showing the variables for the off-diagonal parton 
distribution $H (x, \xi)$ where $x_{1,2} = x \pm \xi$.}
\label{fig:var}
\end{center}
\end{figure}

Moreover it has been proved, using dispersion relations~\cite{DI},
that only distributions in the space-like region $|x|>\xi$ are needed
to describe these processes. The variables are defined in
Fig.~\ref{fig:var}. 

There are insufficient experimental data to determine the skewed parton distributions
with an accuracy comparable to that of the global parton analysis of the conventional
(diagonal) distributions. Fortunately, for small $\xi$, the skewed
distributions can be computed accurately just from the knowledge of
the known integrated conventional distributions. At first sight such a
simplification looks surprising. On the other hand, we know that the
anomalous dimensions which describe the evolution of the Gegenbauer
moments, $G_N$, of the skewed distributions are equal to the
corresponding anomalous dimensions of the conventional Mellin moments,
$M_N$ \cite{Ohrndorf:1981qv,Bukhvostov:1985rn}. This is a consequence
of conformal invariance of the evolution equations. Strictly speaking,
conformal invariance is only valid at leading order (LO). Already at
NLO \cite{Brodsky:1984xk,Mueller:1993hg}, it is violated by the
running of $\alpha_S$, leading to a mixture of the operators at adjacent
orders. But let us start with LO. 

At LO we have equality of the anomalous dimensions of $G_N$ and
$M_N$. Moreover, due to the polynomial property \cite{Ji:1998pc,Ji:1996ek,Ji:1996nm}, 
\begin{equation}
 G_N~=~\sum_{n=0}^N c^N_n \xi^{2n},
\end{equation}
we have $c^N_0=M_N$. That is, from the conventional global parton
analyses we can determine all the Gegenbauer moments of the skewed
parton distributions at small $\xi$ with an accuracy of
$O(\xi^2)$. Then it is simply a technical problem to calculate the $x$
distribution of the skewed partons from the known moments. This
mathematical problem is solved by the Shuvaev transform~\cite{Shuvaev:1999fm}
(which we give explicitly later). Note that at $\xi=0$, in the
diagonal case, there is no mixture of the Mellin moments during the
evolution.  Thus for the skewed distributions such mixing must also
vanish at $\xi=0$. Indeed the mixture of the different operators, due
to violation of conformal invariance\footnote{This violation occurs
  due to the dependence of $\alpha_s$ on the dimensionful parameter
$\Lambda_{\rm QCD}$. It is essentially trivial. In principle, it could
be accounted for by using the `correct' argument of the QCD coupling
in the uppermost cell of Fig.~\ref{fig:var}; namely $\alpha_s(k_t^2)$ and not
$\alpha_s(Q^2)$.}, is of $O(\alpha_s\xi)$
\cite{Brodsky:1984xk,Mueller:1993hg,Belitsky:1998uk,Belitsky:1998gc}. Thus
the Shuvaev transform can be used at NLO with accuracy $O(\xi)$, which
is sufficient for the description of all diffractive processes of
interest. 

This procedure has been called into doubt~\cite{Diehl:2007zu}. First,
we discuss the reason for the doubt and then explain why it is not a
problem in practice. The apparent problem is that to obtain the $x$ distributions
from the moments $G_N$ we must analytically continue the moments into
the complex $N$-plane. If there is a singularity in the right-half
plane, then it may generate a non-negligible correction of $O(\xi/x)$,
instead of the $O(\xi^2)$ correction which came from the difference
between the Gegenbauer and Mellin moments. This large $O(\xi/x)$
correction would destroy the practical use of the Shuvaev
transform. So to justify the Shuvaev transform we require the absence
of singularities in the right-half $N$ plane. 

Now, an arbitrary singularity in the right-half plane will violate the
polynomial property, and so at first sight the danger is
removed. However, it has been shown by
Radyushkin~\cite{Radyushkin:1997ki,Radyushkin:1998es,dd,Musatov:1999xp}
that it 
is possible to form `double distributions' which are not identical to
the Shuvaev transform, but which still satisfy the polynomial
property. The non-polynomial contributions, generated by different
singularities of these double distributions in the right-half plane,
compensate each other to guarantee polynomiality. So the danger
remains. On the other hand, there is no singularity in the right-half
plane in the anomalous dimensions which describe the $q^2$ evolution
of the Gegenbauer moments. Hence the extra singularities must come
only from the input distribution. Now, it is natural to describe the
input distribution at low $x$ in terms of the Regge approach, which is
successful in the description of high energy interactions at low
scales where the conventional (collinear) DGLAP evolution for the skewed
distributions starts. In the Regge approach there are no singularities
in the right-half plane ($j>1$) in the space-like $(x>\xi)$
domain.

\begin{figure} [t]
\begin{center}
\includegraphics[width=0.5\textwidth]{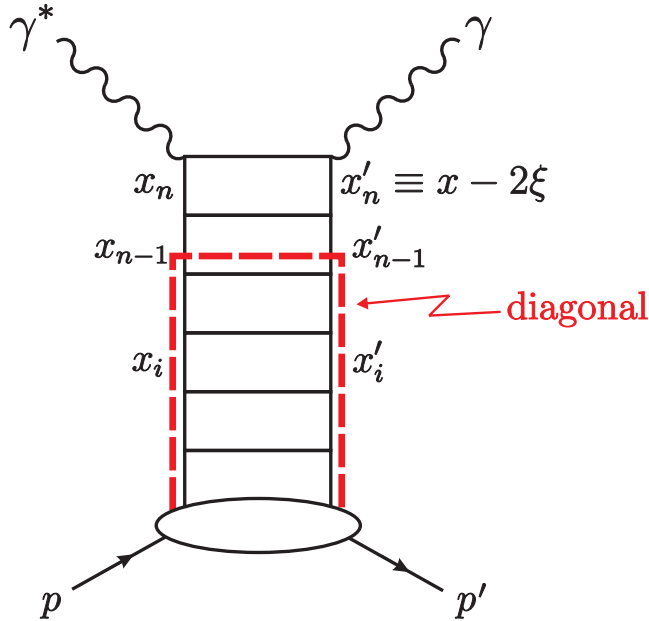}
%\vspace{-3cm}
\caption{Schematic diagram describing deeply virtual Compton scattering (DVCS) in terms of generalised parton distributions. The skewing (or $\xi$ dependence) originates from the uppermost cell; the part of the diagram enclosed by the dashed line has diagonal form due to the strong ordering of $x$ in the Regge limit. This diagram makes clear why, in the low $x$ region, generalised parton distributions may be computed directly from the well known `global' diagonal distributions.}
\label{fig:ord}
\end{center}
\end{figure}
Indeed, in the Regge limit we are
  concerned with the leading log$(1/x)$ summation. Hence, in Fig.~\ref{fig:ord} we have the strong ordering
\begin{equation}
x_i\gg x'_{i+1}, ~~~~~{\rm that ~is}~~|x_i-x'_i| \equiv 2\xi \ll x_i~~~{\rm for}~~~i<n.
\end{equation}
The inequality $\xi \ll x_i$ means that the lower part of the diagram is described by the diagonal distribution. The only $\xi$ dependence comes from the uppermost $x_n$ cell. All other cells have $x_i \gg \xi$. Now, the uppermost cell
  already satisfies conformal invariance. Moreover, at NLO level, this uppermost cell is calculated exactly in terms of the NLO coefficient functions.  Therefore there is no opportunity for a new singularity in the right-half $j>1$ plane.

 With the above physically motivated conjecture, the
Shuvaev transform survives to be of practical value. We emphasise that
the conjecture, of the absence of additional singularities in the
right-half plane, does not follow from first principles or symmetry
arguments, but rather from the assumption of the Regge form of the low
$x$ input. 

Note that in the time-like region, that is at $|x|<\xi$,
we have such singularities. They correspond to the wave functions of
resonances with large spin. Indeed one may add to the distribution,
given by the Shuvaev transform, an additional function with support
only on the interval $|x|<\xi$. However there is no such contribution
in the calculation of the skewed partons via the Shuvaev transform in
the space-like domain.\footnote{As was pointed out in
  \cite{Noritzsch:2000pr}, such `meson-wave-function-like'
  contributions lead to extra terms in the inverse Shuvaev transform,
  which we do not use here. Also note that similar extra terms in the
  effective diagonal functions as proposed in
  \cite{Noritzsch:2000pr} are irrelevant for our case.} 

Another argument in favour of the Shuvaev prescription is the good fit
of the DVCS HERA data obtained at NLO and NNLO in \cite{km}. This fit
was based on the approximation that there is only one pole. 
Recall also the model \cite{Noritzsch:2000pr} based on the Shuvaev
transform and the successful description of the DVCS HERA data in the
dipole model \cite{Favart:2007zz} where the ratio, $R_g$, of the
skewed to diagonal gluon distribution given by the Shuvaev transform
was used.

In \cite{Mueller:2005ed} the Shuvaev transform was criticised to be
``unpractical'' or ``too complicated'', but see also the original 
discussion in~\cite{Polyakov:2002wz}. Here we present and describe a
package that allows for the simple computation of the skewed
distributions at small $x, \xi$ based on the Shuvaev transform.

\section{Description of the Shuvaev transform}
The generalised parton distributions (GPDs) are denoted
\cite{Ji:1998pc,Ji:1996ek,Ji:1996nm} by $H_i(x,\xi)$, $i=q,g$ where
$\xi$ is the skewing parameter, with $x_{1,2}=x\pm\xi$ as defined in
Fig.~\ref{fig:var}.  In fact $H(x,\xi)\equiv H(x,\xi,t,\mu^2)$ for
partons emitted and absorbed at scale $\mu^2$, with the momentum
transfer $t=(p-p^\prime)^2$, and $-1<x<1$.  The values of $t$ and $\xi$ do
not change as we evolve in $\mu^2$.  In the limit $\xi\to 0$ the
skewed distributions will reduce to the usual diagonal partons: 
\begin{eqnarray}
\label{eq:aa}
H_q (x, 0) & = & \left \{ \begin{array}{c} q (x) \quad\quad \;\; {\rm for} \quad x > 0 \\
- \bar{q} (- x) \quad {\rm for} \quad x < 0 \end{array} \right . \,,\nonumber \\
& & \\
H_g (x, 0) & = & x \: g (x). \nonumber
\end{eqnarray}

We briefly discuss the double-integral (that is, the Shuvaev
transform) used to calculate the skewed distributions for $\xi \ll 1$
in the space-like region $|x|>\xi$. 
First we introduce an auxiliary function $f_\xi(x,t)$, whose Mellin moments
are equal up to normalisation to the Gegenbauer moments of the skewed
function.  The function formed from the DGLAP evolution of the
auxiliary function is referred to as the `effective diagonal
function', $f(x^\prime)$.  Neglecting the $t$ dependence, we must find the
kernel $K(x,\xi;x^\prime)$ relating $f(x^\prime)$ to the skewed distribution, 
\begin{equation}
\label{Kqq}
H(x,\xi)\,=\,\int {\rm d}x^\prime K(x,\xi;x^\prime)\,f(x^\prime).
\end{equation}
It is convenient to first weaken the singularity in the $x^\prime$ integral
by an integration by parts. With this it can be shown
that~\cite{Shuvaev:1999ce} 
\begin{equation}
\label{eq:shuvq}
H_q (x, \xi) \; = \; \int_{-1}^1 \: {\rm d}x^\prime \left [ \frac{2}{\pi} \: {\rm Im} \: \int_0^1 
\: \frac{{\rm d}s}{y (s) \: \sqrt{1 - y(s) x^\prime}} \right ] \:
\frac{{\rm d}}{{\rm d}x^\prime} \left ( 
\frac{q (x^\prime)}{| x^\prime |} \right ),
\end{equation}
\begin{equation}
\label{eq:shuvg}
H_g (x, \xi) \; = \; \int_{-1}^1 \: {\rm d}x^\prime \left [ \frac{2}{\pi} 
\: {\rm Im} \: \int_0^1 \: \frac{{\rm d}s (x + \xi (1 - 2s))}{y (s) \: \sqrt{1 - y (s) x^\prime}} 
\right ] \: \frac{{\rm d}}{{\rm d}x^\prime} \left ( \frac{g (x^\prime)}{| x^\prime |} \right ) \:,
\end{equation}
\begin{equation}
\label{eq:a200}
y(s)=\frac{4s(1-s)}{x+\xi(1-2s)}\,.
\end{equation}
Equations (\ref{eq:shuvq}) and (\ref{eq:shuvg}) are (up to a variable
substitution $z = 1/(x^{\prime}y(s))$) the form used
to compute the Shuvaev transform in the package decribed below.

Incidentally, it is also possible to solve the inverse problem. That
is to obtain the diagonal distribution from a known skewed
distribution $H(x,\xi)$ at a given value of $\xi$. Although this is
not needed for our discussion, for completeness, we give the details
in the Appendix. 

Note that the transforms, (\ref{eq:shuvq}) and (\ref{eq:shuvg}), say
nothing about the $t$ (or $p_T$) dependence of the GPDs. Strictly
speaking they are written for $p_T=0$. Usually the factorisation 
\begin{equation}
 H(x,\xi;p_T)~=~H(x,\xi)~F(p_T)
\end{equation}
is assumed, where $F(p_T)$ is just the proton form factor.

\section{Approximate predictions of GPDs for small $x$ and $\xi$}
We see that (\ref{eq:shuvq}) and (\ref{eq:shuvg}) determine the
behaviour of the skewed distributions in the small $x, \xi$ domain
entirely in terms of the diagonal distributions. Before we perform the
exact evaluation of these integral expressions for the GPDs, it is
informative to recall that approximate expressions can be obtained by
making the physically reasonable small $x$ assumption that the
diagonal partons are given by 
\begin{equation}
x q (x) \; = \; N_q \: x^{- \lambda_q}, \quad\quad xg (x) \; = \; N_g \: x^{- \lambda_g}.
\label{eq:pdfpower}
\end{equation}
Then we can perform the $x^\prime$ integration
analytically.\footnote{With the substitution $z = 1/x^{\prime}
  y(s)$ one can then use $$ \int_0^1 \: {\rm d}z \: z^{\lambda +
    \frac{3}{2}} \: (1 - z)^{- \frac{1}{2}} \; = \; \frac{\Gamma \left
      ( \lambda + \frac{5}{2} \right ) \: \Gamma \left ( \frac{1}{2}
    \right )}{\Gamma (\lambda + 3)}\,,$$where we have set the lower limit to
  zero. We have checked numerically that this is a very good
  approximation, with accuracy of the order of $10^{-4}$ for $\lambda
  \simeq 0.2$ in the small $x$ region. (That this approximation is good
  for small $\xi$ is in line with the findings
  of~\cite{Noritzsch:2000pr}.) } We obtain
\be
\label{eq:shuvappr}
H_i (x, \xi; t) \; = \; N_i \: \frac{\Gamma \left ( \lambda + \frac{5}{2} \right 
)}{\Gamma (\lambda + 2)} \: \frac{2}{\sqrt{\pi}} \: \int_0^1 \: {\rm d}s \left [ x 
+ \xi (1 - 2s) \right ]^p \: \left [ \frac{4s (1 - s)}{x + \xi (1 - 2s)} \right ]^{\lambda_i + 
1} \: G (t)
\ee
with $i = q$ or $g$, and where $p = 0$ and 1 for quarks and the gluon
respectively. 

At first sight it appears that for singlet quarks (where $\lambda_q >
0$ and $p = 0$) we face a strong singularity in integral
(\ref{eq:shuvappr}) when the term $D \equiv x + \xi (1 - 2s)
\rightarrow 0$ in the denominator. Fortunately the singlet quark
distribution is antisymmetric in $x$. To obtain the imaginary part of
the integral (\ref{eq:shuvq}) we must choose $x^\prime > 0$ for $D >
0$ and $x^\prime < 0$ for $D < 0$. Therefore we must treat
(\ref{eq:shuvappr}) as a principal value integral and take the
difference between the $D \rightarrow 0+$ and $D \rightarrow 0-$
limits. Thus the main singularity is cancelled and (\ref{eq:shuvappr})
becomes integrable for any $\lambda_q < 1$.

Note that the dominant contribution to the $x^\prime$ integrations of
(\ref{eq:shuvq}) and (\ref{eq:shuvg}) comes from the region of small
$x^\prime \sim x, \xi$. Indeed with the input given by (\ref{eq:pdfpower}),
the integral for the quark distribution has a strong singularity at
small $x^\prime$
\be
\label{eq:b21}
I_q \; \sim \; \int \: {\rm d}x^\prime (x^\prime)^{- \lambda_q - 3} \: {\rm Im} \left ( 
\frac{1}{y (s) \sqrt{1 - y (s) x^\prime}} \right ).
\ee
However when we take the imaginary part, the $x^\prime$ integration is
cut-off by the theta function $\theta (x^\prime - 1/y (s))$ at
\be
\label{eq:c21}
x^\prime \; = \; 1/y (s) \; \sim \; x + \xi (1 - 2s),
\ee
which implies that $x'$ is always greater than $x/2$.  So we obtain the small $\xi$ behaviour $I_q \sim \xi^{- \lambda_q -
  1}$, and the distribution (\ref{eq:shuvq}) has the form
\be
\label{eq:d21}
H_q (x, \xi) \; = \; \xi^{- \lambda_q - 1} \: F_q (x/\xi).
\ee
Similarly it follows that $H_g = \xi^{- \lambda_g} F_g (x/\xi)$.

It is illuminating to evaluate the ratio $R$ of the GPD to its
diagonal parent distribution; that is 
\be
\label{eq:a22}
R \; = \; \frac{H (x, \xi)}{H (x + \xi, 0)},
\ee
where the only free parameter is $\lambda$, the exponent which fixes
the $x^{-\lambda}$ behaviour of the input diagonal partons, as in
(\ref{eq:pdfpower}). Notice that on account of (\ref{eq:d21}) the ratios
$R$ at small $x$ and $\xi$ are a function of only the ratio of the
variables $x/\xi$. 

\begin{figure}
\begin{center}
\includegraphics[width=0.7\textwidth]{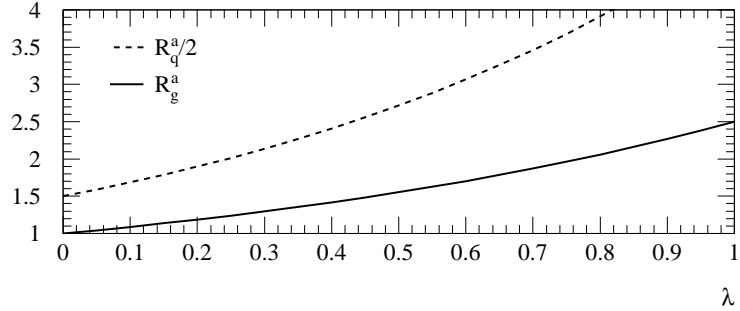}
\vspace{-7mm}
\caption{The off-diagonal to diagonal ratio, $R^a$, at small $x=\xi$
  versus the power $\lambda$ which specifies the $x^{-\lambda}$
  behaviour of the input diagonal parton as in (\ref{eq:pdfpower}). Note
  that the quark singlet ratio has been divided by two.} 
\label{fig:R}
\end{center}
\end{figure}
Assuming the pure power behaviour of the partons, the ratios at $x =
\xi$ are given explicitly in analytic form as
\begin{equation}
R^a \; = \; \frac{H (\xi, \xi)}{ H (2 \xi, 0)}\; =\; \frac{2^{2\lambda+3}}{\sqrt{\pi}}
\frac{\Gamma(\lambda + 5/2)}{\Gamma(\lambda + 3 + p)}\, ,
\label{eq:ranalytic}
\end{equation}
where $p = 0$ for quarks and $p = 1$ for gluons. These ratios are
plotted in Fig.~\ref{fig:R} as a function of $\lambda$. We see that
the off-diagonal or `skewed' effect (the ratio $R^a$) is much
stronger for singlet quarks than for gluons. The explanation is
straightforward. At low $x$ the distributions are driven by the double
leading logarithmic evolution of the gluon distribution. At each step
of the evolution the momentum fractions $x_i$ are strongly ordered
($x_1^\prime \gg x_1, x_2^\prime \gg x_2$ on Fig.~1). For gluons it is
just the \lq\lq last splitting function\rq\rq \ $P_{gg} (x_2,
x_2^\prime; \xi)$ which generates the main $\xi$ dependence, or
skewedness, of the distribution. However for the sea or singlet quarks
it is necessary to produce a quark with the help of $P_{qg}$ at the
last splitting. The splitting function $P_{qg}$ has no logarithmic
$1/z = x_2^\prime/x_2$ singularity and so $x_2$ is the order of
$x_2^\prime$. Consequently both the splitting functions $P_{qg} (x_2,
x_2^\prime; \xi)$ and $P_{gg} (x_2^\prime, x_2^{\prime\prime}; \xi)$
generate the asymmetry of the off-diagonal distribution. Hence, at low
$x$, the singlet quark has a much stronger off-diagonal effect than
the gluon. 

We emphasise that the analytical expression for the ratio $R^a$,
(\ref{eq:ranalytic}), is valid in the limit $x=\xi$ only and
that it assumes a pure power behaviour of the diagonal partons,
(\ref{eq:pdfpower}). As will be discussed below, the latter is a
good approximation for global fit partons in the small $x$
regime. However, for $x > \xi$ the difference between the result of the
complete (double integral) Shuvaev transform, (\ref{eq:shuvq},
\ref{eq:shuvg}), and the approximate analytic formula for $R^a$, (\ref{eq:ranalytic}), 
based on the limit $x=\xi$, is quite large. This is evident from Fig.~\ref{fig:rcompquarksgluons} below.

\section{Evaluation of the GPDs using the Shuvaev transform}
\begin{figure}
\begin{center}
\includegraphics[width=0.7\textwidth]{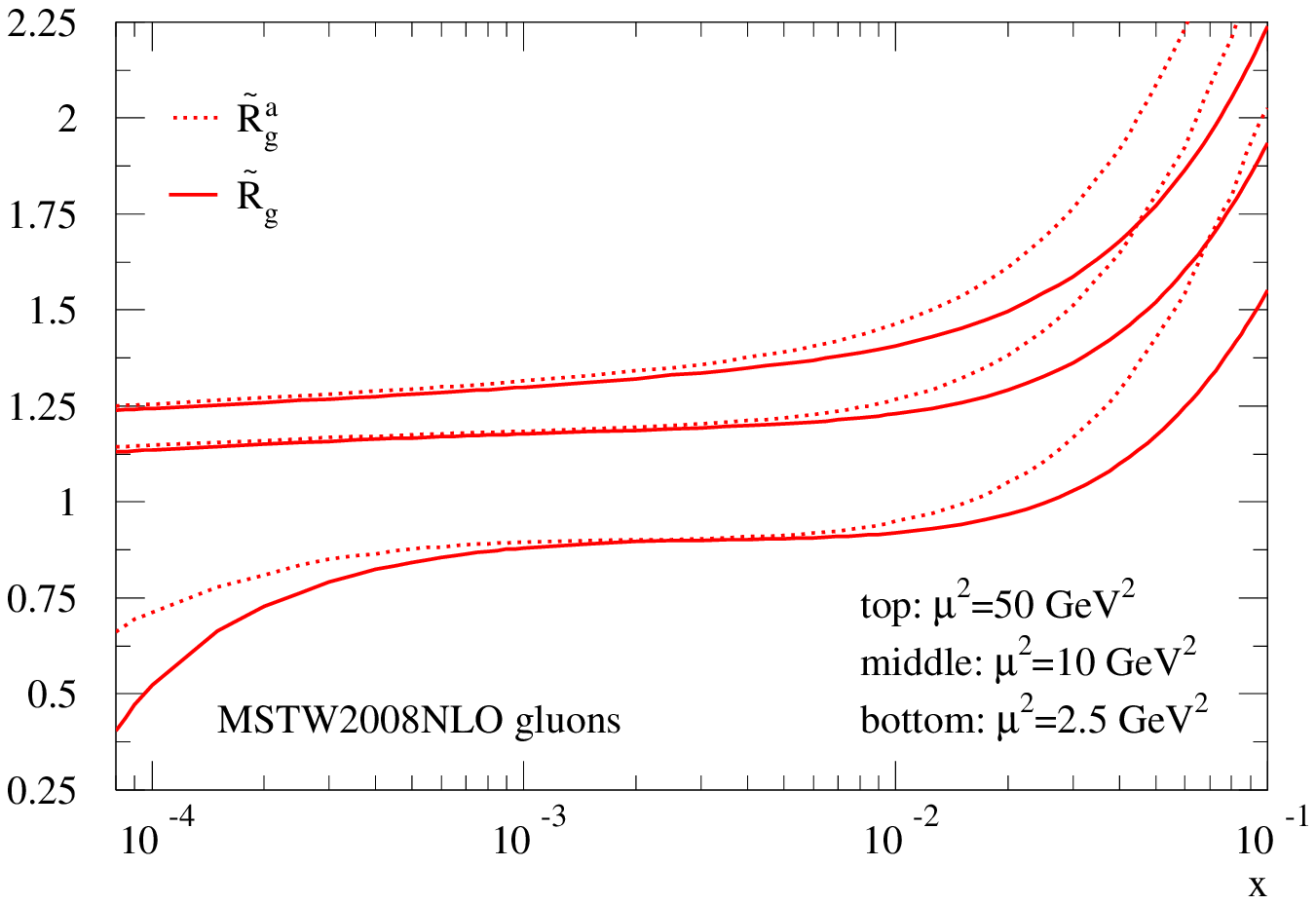}\\[-2cm]
\includegraphics[width=0.7\textwidth]{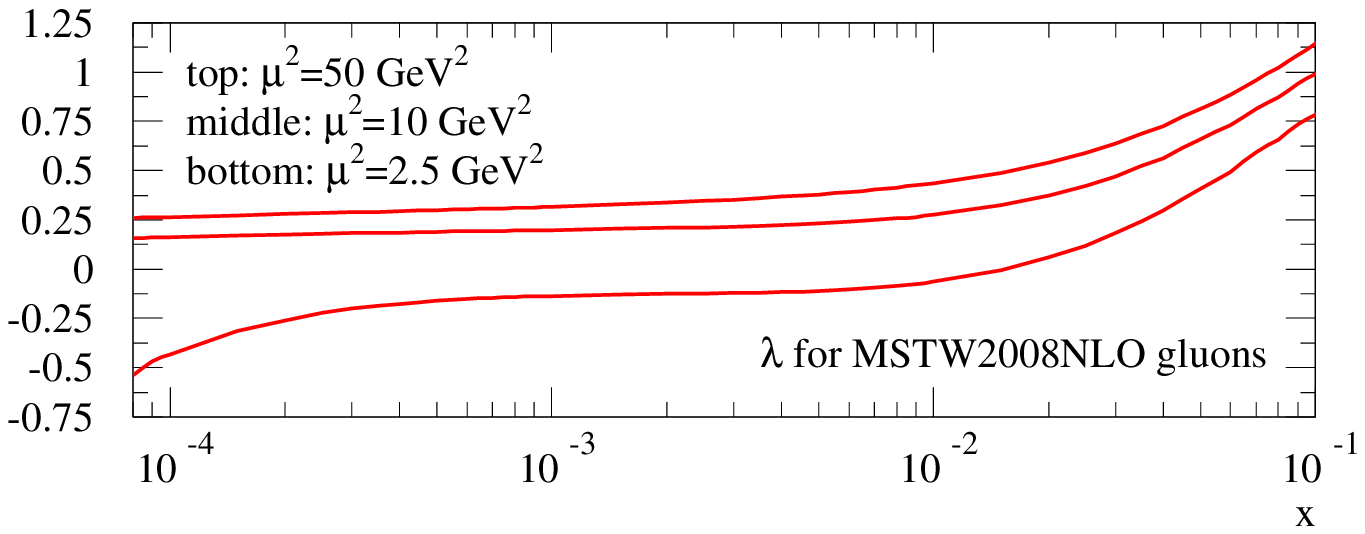}
\vspace{-8mm}
\caption{Lower panel: Values for the effective power
  $\lambda$ (evaluated at $x$) of the MSTW2008NLO gluon for the three scales
  $\mu^2=2.5,\,10,\,50\,\mathrm{GeV}^2$. Upper panel: Analytic
  approximation $\tilde R^a_g$ (dotted lines) as defined in
  (\protect{\ref{eq:Rtildeanalytic}}) compared to the ratio $\tilde R_g$
  (solid lines) as defined in (\protect{\ref{eq:Rtilde}}) for the full
  Shuvaev transform.} 
\label{fig:rlambdagluon}
\end{center}
\end{figure}
\begin{figure}
\begin{center}
\includegraphics[width=0.7\textwidth]{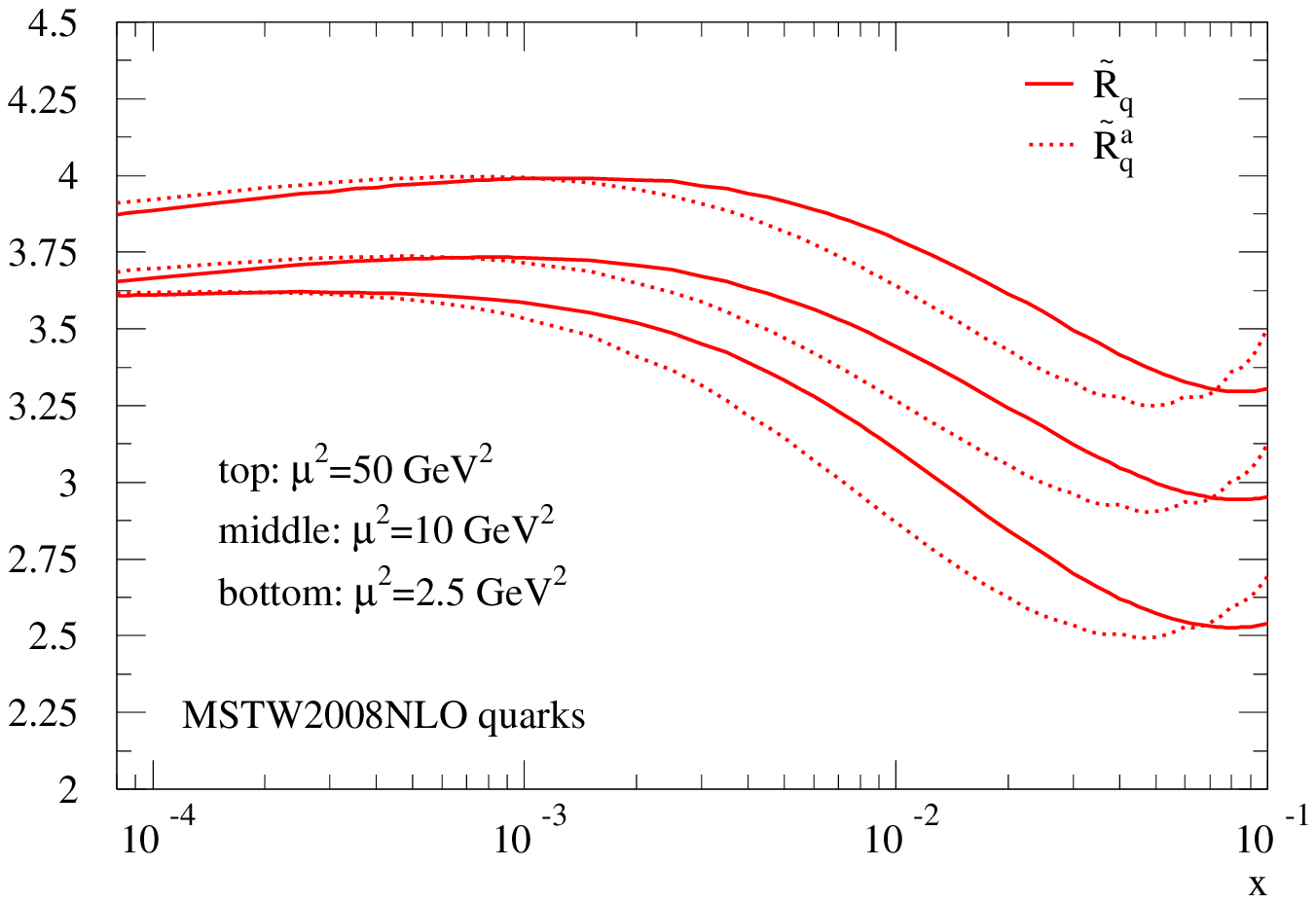}\\[-2cm]
\includegraphics[width=0.7\textwidth]{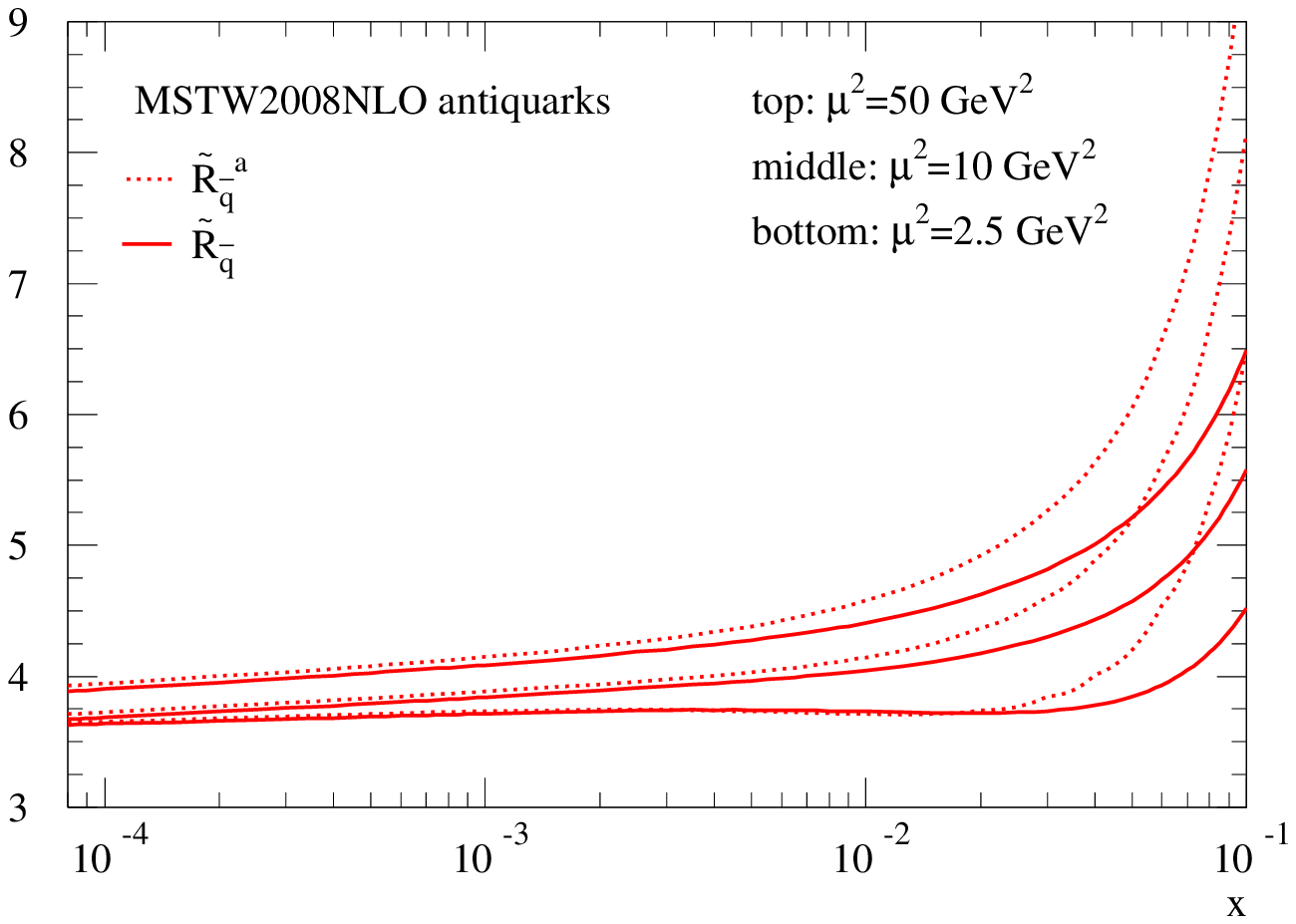}\\[-2cm]
\includegraphics[width=0.7\textwidth]{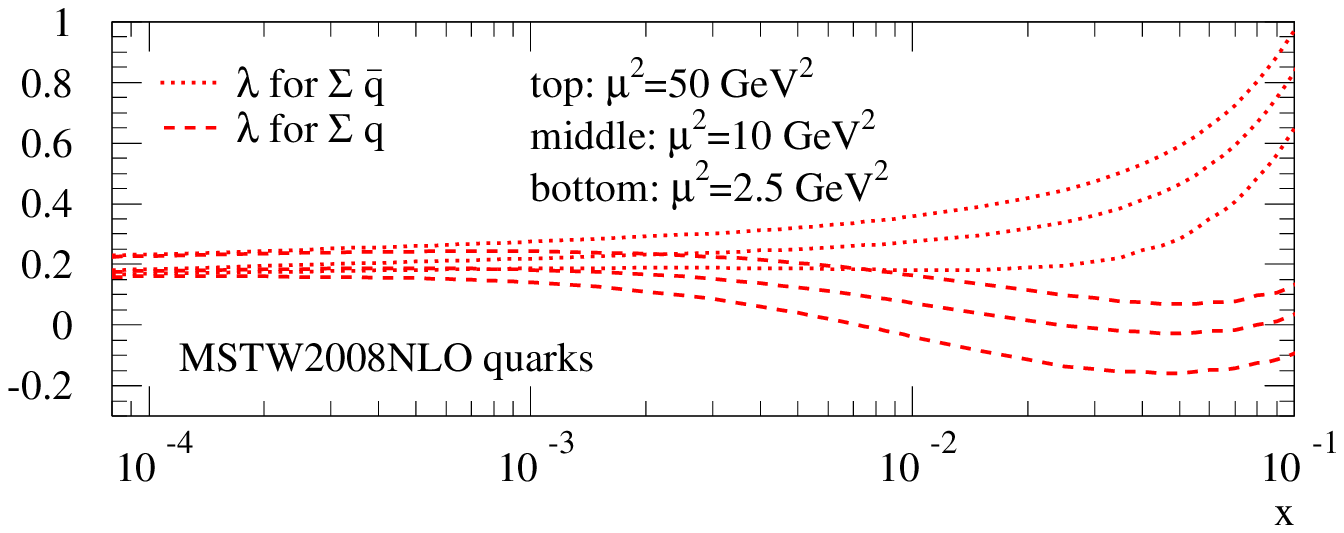}
\vspace{-8mm}
\caption{Lower panel: Values for the effective power
  $\lambda$ (evaluated at $x$) for the sum of the MSTW2008NLO $u, d, s$
  quarks (dashed) and antiquarks (dotted lines) for the three scales
  $\mu^2=2.5,\,10,\,50\,\mathrm{GeV}^2$. Upper two panels: $\tilde
  R^a_q$ as defined in (\ref{eq:Rtildeanalytic}) shown as dotted lines
  for quarks (top panel) and antiquarks (middle panel) compared to the
  ratio $\tilde R_q$ for the full Shuvaev transform as defined from
  (\protect{\ref{eq:Rtilde}}) (solid lines).} 
\label{fig:rlamdaquarks}
\end{center}
\end{figure}
In the above section we have obtained an approximate determination of
the GPDs valid for $x \simeq \xi$. We now perform a precise evaluation
of the Shuvaev transforms of (\ref{eq:shuvq}) and (\ref{eq:shuvg}),
valid for all small $x$, $\xi$, and compare our results to the
approximation based on (\ref{eq:shuvappr}). This is done in
Fig.~\ref{fig:rlambdagluon} for the gluon and in
Fig.~\ref{fig:rlamdaquarks} for the sums of $u, d, s$
quarks and antiquarks, using MSTW2008NLO~\cite{mstw2008} (conventional
global fit) parton distributions as input. Thus our GPDs correspond to
using the $\overline{\rm MS}$ renormalisation scheme and the General
Mass Variable Flavour Number Scheme adopted by MSTW. 

\begin{figure}
\begin{center}
\includegraphics[width=0.49\textwidth,bb=25 5 410 360]{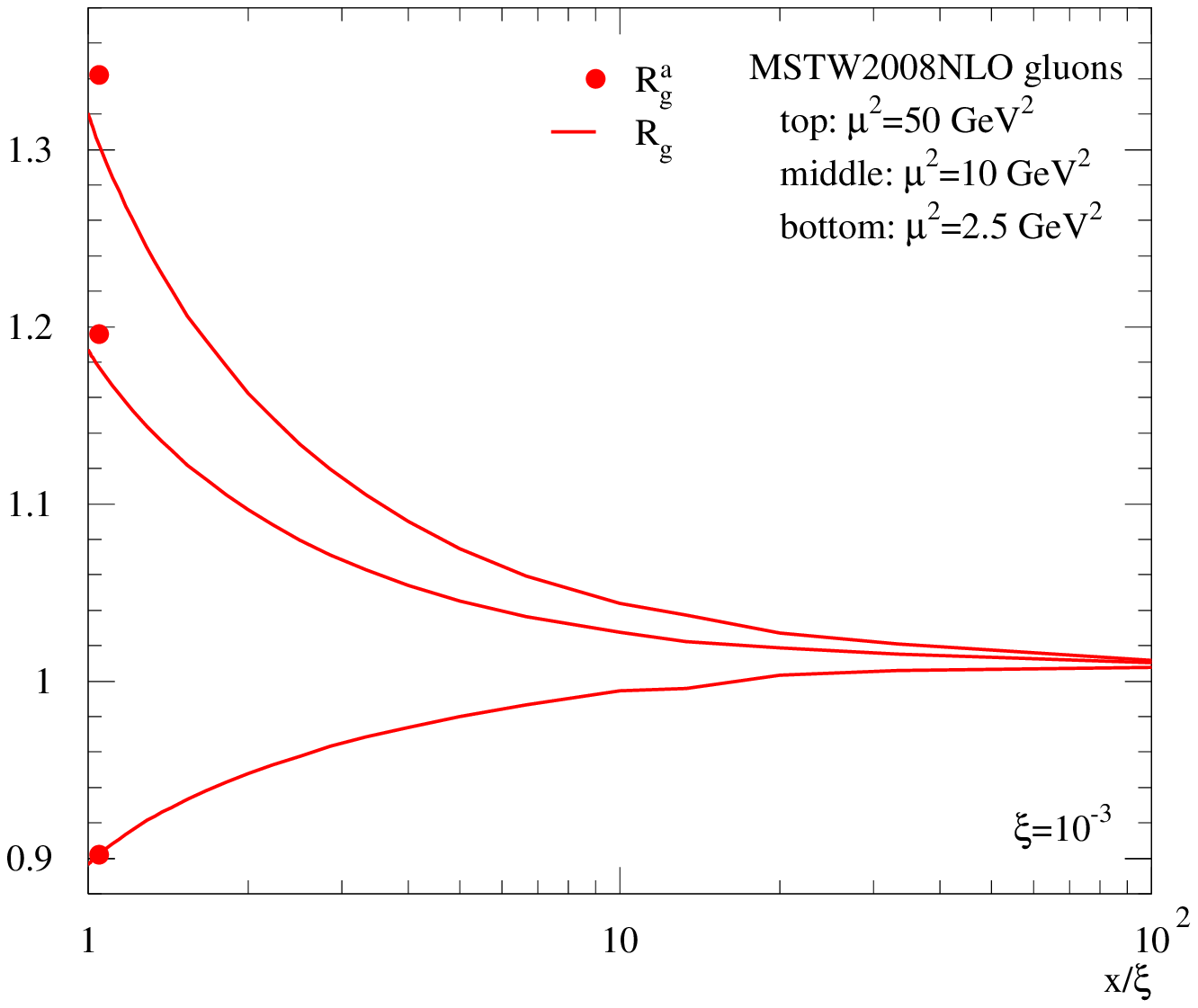}
\includegraphics[width=0.49\textwidth,bb=25 5 410 360]{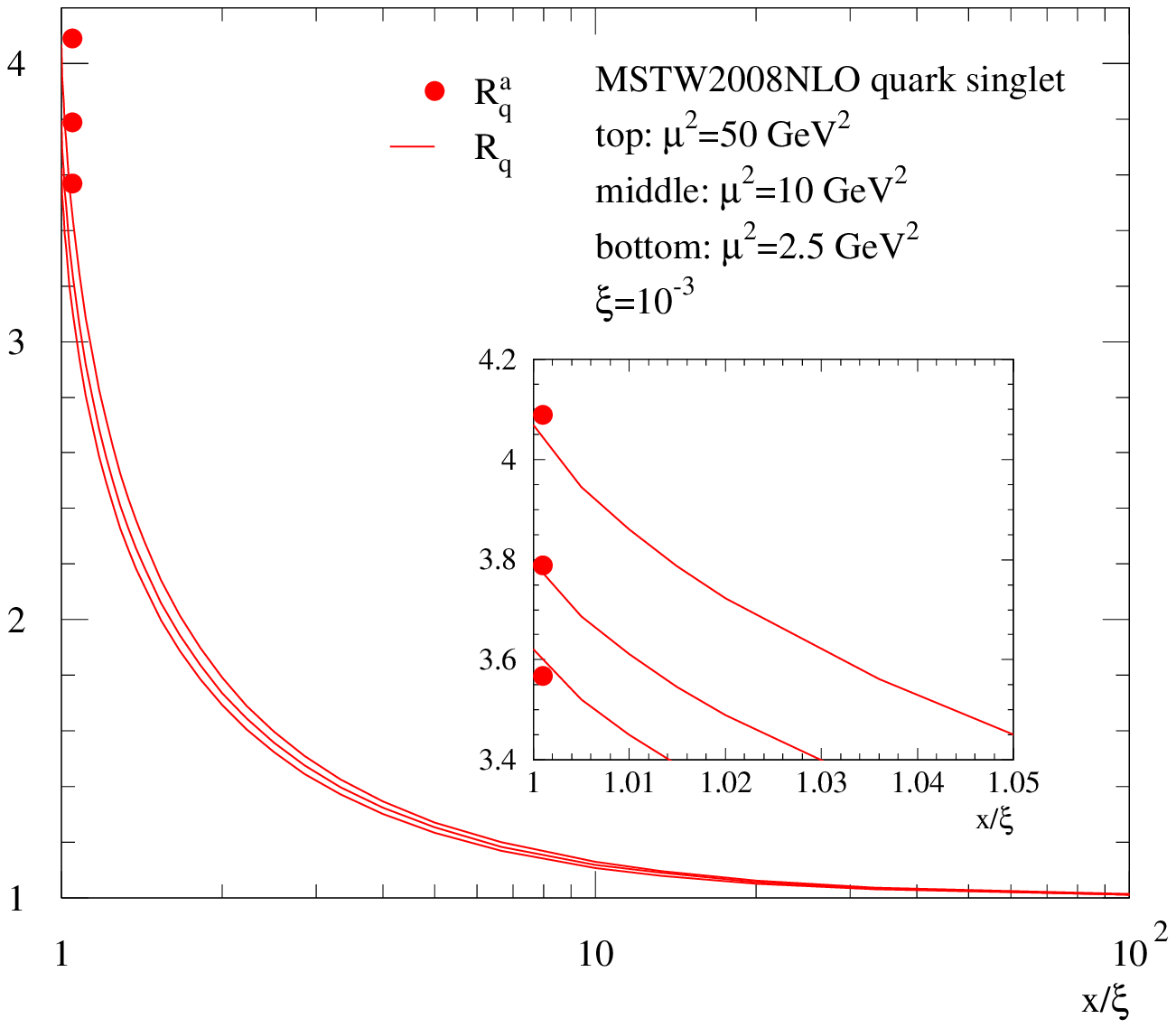}
\vspace{-5mm}
\caption{Left panel: Analytic $R^a_g$ from (\ref{eq:ranalytic}) (fat dots)
  compared to the ratio $R_g$ from (\ref{eq:a22}) (solid lines) using the full Shuvaev
  transform (\ref{eq:shuvg}), at the three scales
  $\mu^2=2.5,\,10,\,50\,\mathrm{GeV}^2$ for the MSTW2008NLO
  gluon and $\xi = 10^{-3}$ as a function of $x/\xi$. Right panel:
  same as left panel but for the sum of $u, d, s$ quarks and
  antiquarks. The insert shows a blow-up of the regime $x \simeq
  \xi$.} 
\label{fig:rcompquarksgluons}
\end{center}
\end{figure}
In the lower panels of the figures the
effective powers $\lambda$ are plotted as a function of
$x$ for three different scales, $\mu^2 = 2.5, 10, 50$ GeV$^2$. In the
upper panels both the ratios 
\begin{equation}
\tilde R \; = \; \frac{H(x/2, x/2)}{H(x, 0)}
\label{eq:Rtilde}
\end{equation} 
for the full Shuvaev transform and the corresponding analytical
approximations 
\begin{equation}
\tilde R^a \; = \; \frac{H (x/2, x/2)}{H(x, 0)} \; = \; \frac{2^{2\lambda+3}}{\sqrt{\pi}}
\frac{\Gamma(\lambda + 5/2)}{\Gamma(\lambda + 3 + p)} \qquad
\mbox{(with $\lambda$ evaluated at $x$)}
\label{eq:Rtildeanalytic}
\end{equation} 
are given, again for the three scales. It is
clear that, depending on the values of $x$ and the scale, the
deviation from a pure power can lead to a sizeable difference between
the analytic approximation and the full result. However, for small $x
\lapproxeq 2\cdot 10^{-3}$ and not too small scales the deviation is
quite small in the case of the global fit partons MSTW2008NLO. 
Actually the difference is smaller than it appears at first sight
in Fig.~\ref{fig:rlamdaquarks} since the vertical scale does not
extend to zero.

\begin{figure}
\begin{center}
\includegraphics[width=0.49\textwidth,bb=25 5 410 360]{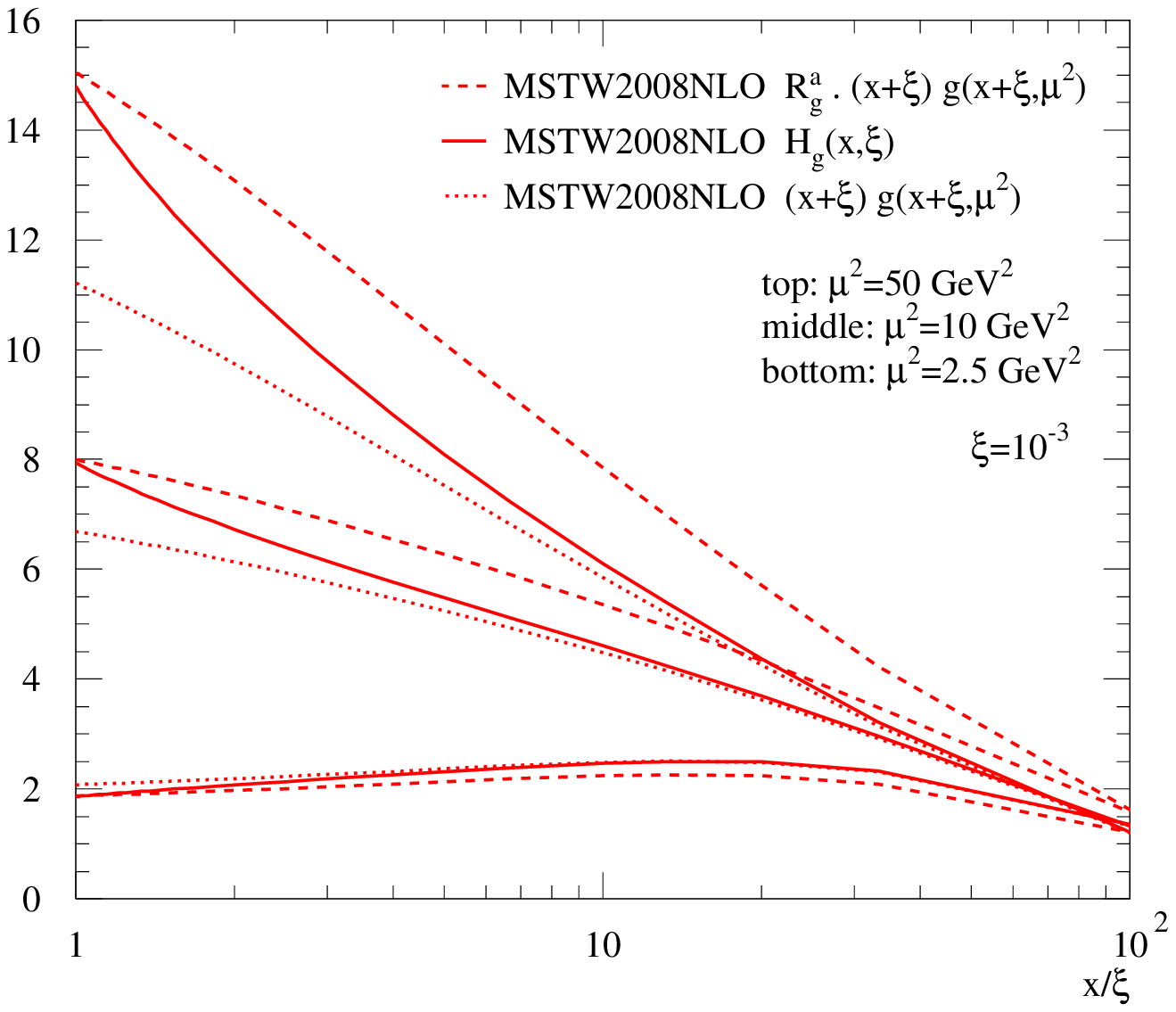}
\includegraphics[width=0.49\textwidth,bb=25 5 410 360]{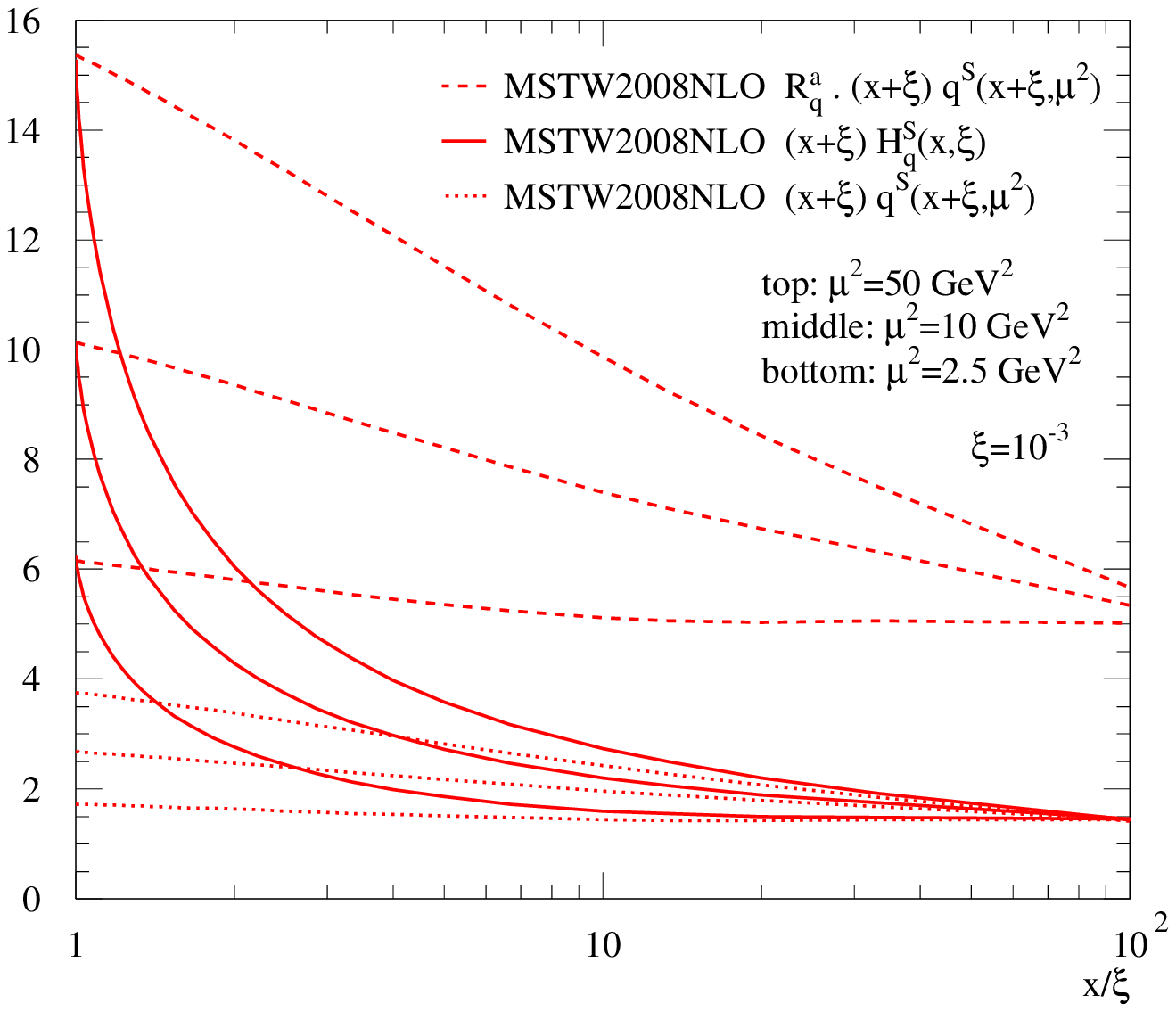}
\vspace{-5mm}
\caption{Comparison of skewed (solid) and diagonal (dotted lines) gluon (left) and
  singlet quark (right panel) MSTW2008NLO distributions for $\xi=10^{-3}$ and
  the three scales $\mu^2=2.5,\,10,\,50\,\mathrm{GeV}^2$ as a function
  of $x/\xi$. Also shown is the product of the diagonal distributions
  with the analytical skewing enhancement factor $R^a$ obtained in the
  $x=\xi$ limit from (\ref{eq:ranalytic}) (dashed lines).} 
\label{fig:compquarksgluons}
\end{center}
\end{figure}
Frequently, skewing corrections are taken into account via simple
multiplication with the ratio $R^a$. One may fear that this
approximation of `maximal skewing' overestimates the real skewing
effect which comes from integrating over the whole possible range $\xi
\leq |x|$. Possible effects are exemplified in
Fig.~\ref{fig:rcompquarksgluons} for MSTW2008NLO gluons (left) and
the sum of $u, d, s$ quarks and antiquarks (right panel). Here the
limit $R^a$ from (\ref{eq:ranalytic}) (fat dots) is compared to the full
ratio $R$  from (\ref{eq:a22}) (solid lines) as a function of
$x/\xi$ for $\xi = 10^{-3}$ at three different scales, $\mu^2 = 2.5,
10, 50$ GeV$^2$. 

\begin{figure}
\begin{center}
\includegraphics[width=0.49\textwidth,bb=25 5 410 360]{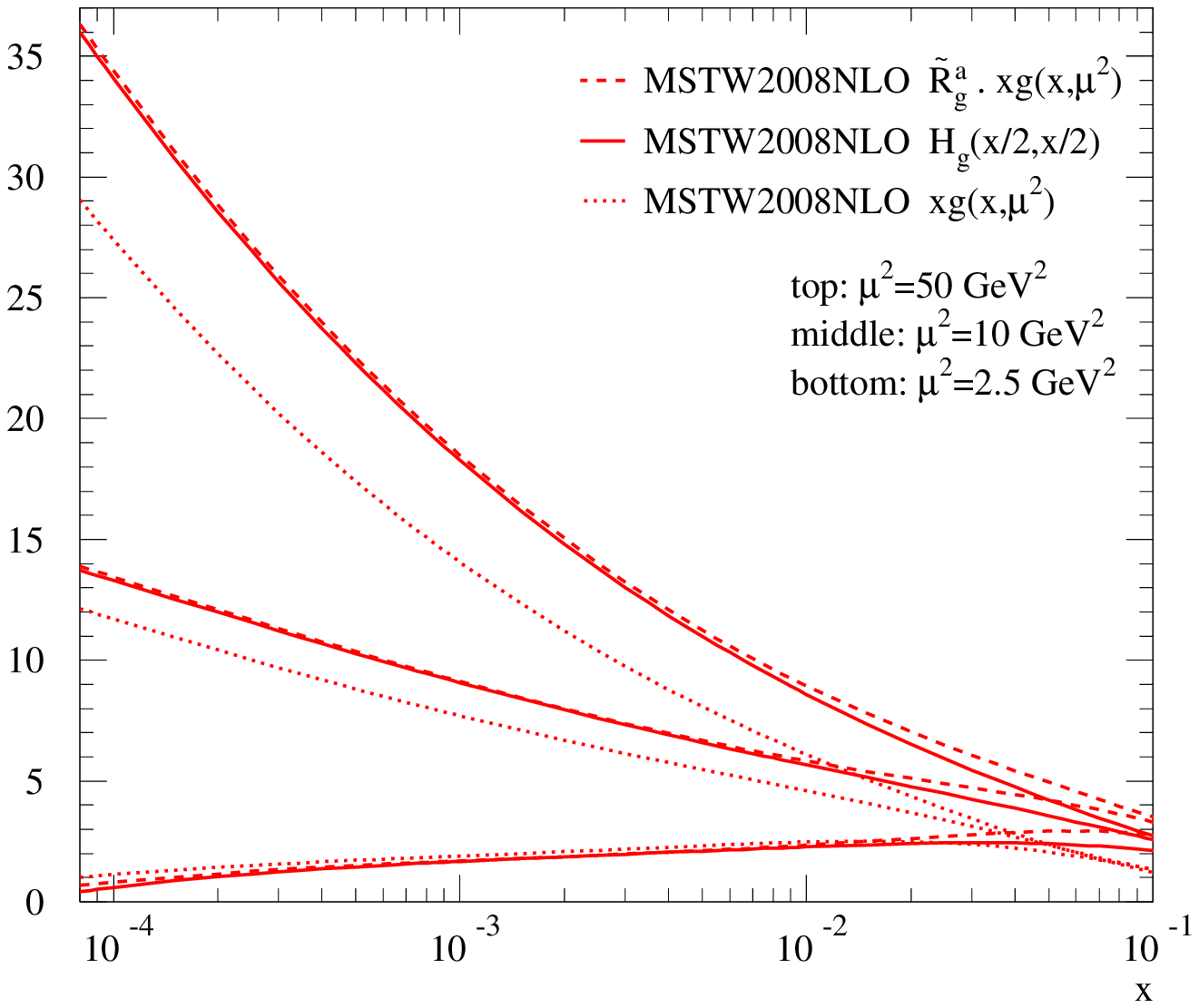}
\includegraphics[width=0.49\textwidth,bb=25 5 410 360]{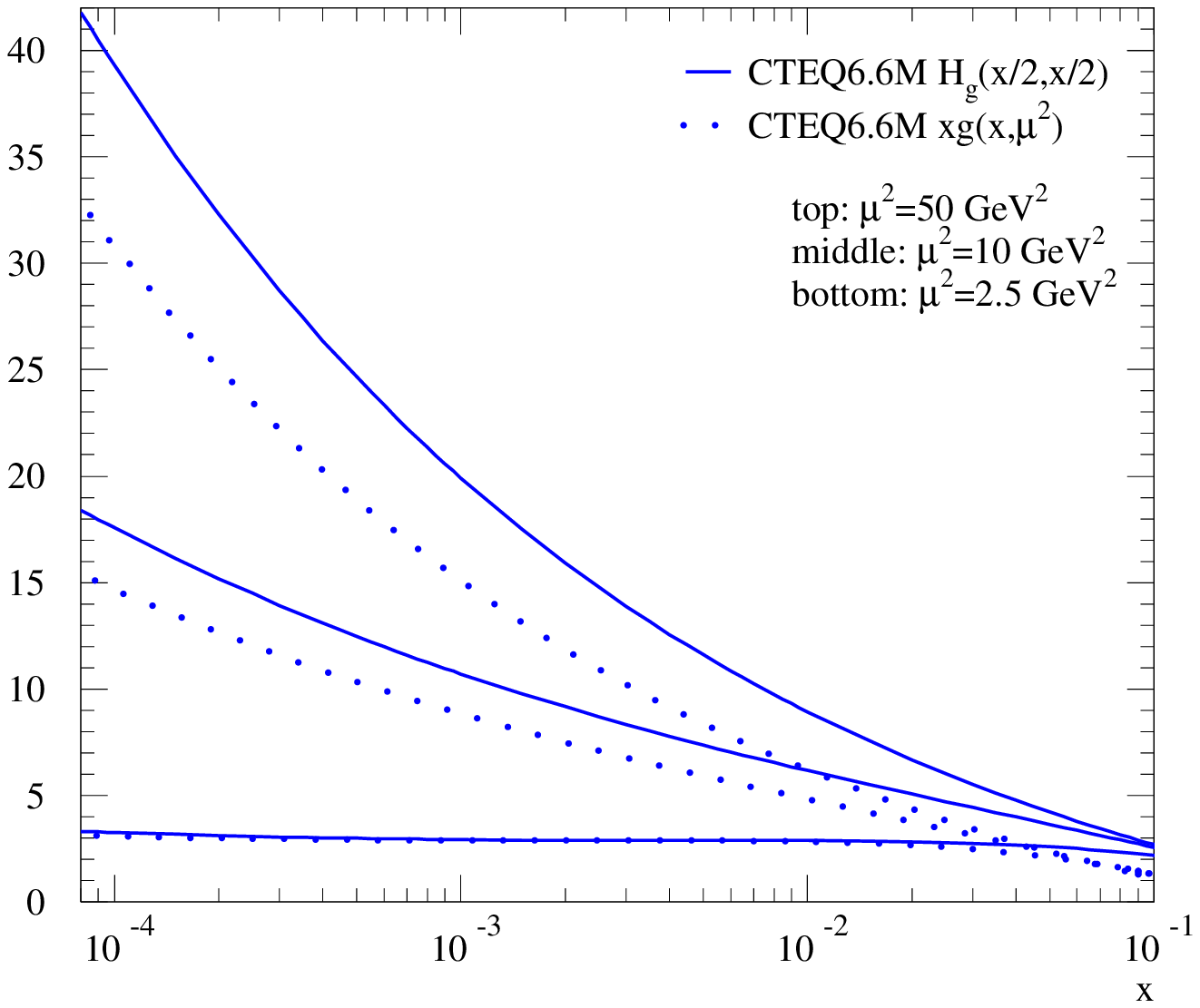}
\vspace{-5mm}
\caption{Skewed (solid and dash-dotted) and diagonal (dotted lines)
  MSTW2008NLO (left) and CTEQ6.6M (right panel) gluon distributions
  at scales $\mu^2=2.5,\,10,\,50\,\mathrm{GeV}^2$. The dashed lines
  show the skewed gluon if the analytic approximation $\tilde R^a$ from
  (\protect{\ref{eq:Rtildeanalytic}}) is used (MSTW only).} 
\label{fig:mrstandcteqgluonsx}
\end{center}
\end{figure}
\begin{figure}
\begin{center}
\includegraphics[width=0.6\textwidth]{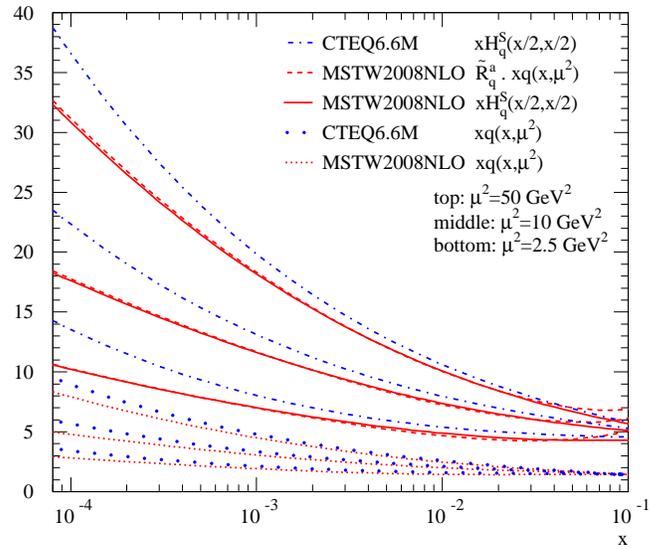}
\vspace{-5mm}
\caption{Skewed (solid) and diagonal (dotted lines) quark singlet
  distributions at scales $\mu^2=2.5,\,10,\,50\,\mathrm{GeV}^2$, for
  MSTW2008NLO and CTEQ6.6M as indicated in the legend. Shown as dashed
  lines (MSTW only) is the approximation using the analytic skewing
  factor $\tilde R^a$ from (\protect{\ref{eq:Rtildeanalytic}}).} 
\label{fig:quarksx}
\end{center}
\end{figure}
In Fig.~\ref{fig:compquarksgluons} a similar comparison is shown for the 
skewed gluon (left) and singlet quark distributions (right panel) as
a function of $x/\xi$ for $\xi = 10^{-3}$ at the three scales. The
solid lines show the results for the full Shuvaev transforms, $H_g(x,
\xi)$ and $(x+\xi) H^s_q(x, \xi)$, whereas the dashed lines are
obtained as the product of the analytical skewing factors $R^a$ (in
the limit $x=\xi$) with the diagonal partons evaluated at $x+\xi$. The
dotted lines show the diagonal MSTW2008NLO partons for comparison.
It is clear from Figs.~\ref{fig:rcompquarksgluons} and
\ref{fig:compquarksgluons} that, at least in the cases under
consideration, the overestimate of skewing effects through simple
multiplication with $R^a$ could be sizeable, although as expected the
agreement when $x = \xi$ is good.

In Figs.~\ref{fig:mrstandcteqgluonsx} and \ref{fig:quarksx} we finally
show the skewed gluon, $H_g(x/2,x/2)$, and singlet quark
distributions, $x H^S_q(x/2,x/2)$ (solid lines), compared to the
corresponding diagonal partons (dotted lines) as a function of $x$ and
at three scales for both MSTW2008NLO~\cite{mstw2008} and
CTEQ6.6M~\cite{cteq}. For MSTW, the analytical approximation for the
skewing using the factor $\tilde R^a$ is also shown (dashed lines). 

Some features of these results are especially noteworthy. First, we
emphasise that for $x \gg \xi$ the skewed distribution $H(x,\xi)$ becomes close
to the diagonal distribution. This is trivial, since for $x \gg \xi$
we can neglect the $\xi$ dependence. Secondly, we see a much larger
skewing enhancement for the quarks than for the gluons. Indeed, it is
known that, in the leading $\ln 1/x$ approximation (LLA), the skewed
distributions are equal to the diagonal ones \cite{BL}. As mentioned
above, this arises
because, in the LLA, the longitudinal fractions, $x$, of the momentum
are strongly ordered. Already at the first evolution step we have $x
\gg \xi$. Hence the $\xi$ dependence becomes negligible. Now for
gluons, with spin 1, the LLA correponds to a flat $xg(x)=constant$
distribution. With a parameterisation of the form $xg \sim
x^{-\lambda}$, we see from Fig.~\ref{fig:rlambdagluon} that for
$\lambda>0~(\lambda<0)$ the skewed distribution is enhanced
(suppressed) in comparison with the diagonal gluon. 
On the other hand, for a pair of $t$-channel quarks with spin
$\frac{1}{2}$, the LLA corresponds to a flat $q(x)=constant$
behaviour, see (\ref{eq:aa}). That is, for the form $xq\sim
x^{-\lambda}$, we have no skewed effect if $\lambda=-1$, but a large
enhancement for small $\lambda$ close to zero. In the latter case,
with $\lambda \gapproxeq 0$, the structure of the loop integration is
such that it prefers to transfer the major part of the momentum flow
along {\it one} quark propagator. 

This behaviour is demonstrated in Figs.~\ref{fig:rcompquarksgluons}
and \ref{fig:compquarksgluons}, at the lowest scale. We see that the
Shuvaev transform actually {\em suppresses} the gluon. This is
expected when we look at the $x$ behaviour of the diagonal gluon;
inspection of Fig.~\ref{fig:rlambdagluon} shows that $\lambda_g$
becomes {\it negative} here. This is also demonstrated in the
$\mu^2=2.5\,\mathrm{GeV}^2$ curves in
Fig.~\ref{fig:mrstandcteqgluonsx}. For MSTW gluons, we have a
suppression of the diagonal gluon, whereas for CTEQ gluons, there is
no net effect at this particular choice of parameters. This behaviour
is demonstrated with the lowest curve for $x \lapproxeq 3\cdot
10^{-2}$ in the lower panel of Fig.~\ref{fig:mrstandcteqgluonsx}; in
this regime CTEQ partons have $\lambda_g \simeq 0$. As the scale
increases, the transform enhances the diagonal gluons.

We emphasise that the use of the $xg,~xq \propto x^{-\lambda}$ forms does not mean that we assume power-like asymptotics for $x \to 0$. Clearly, this oversimplified parameterisation can be valid only in a limited range of energy or $x$. We have used it here simply to qualitatively illustrate some of the main features of the skewed effect. Actually, in practice, and in the supplied grids, we determine the GPDs directly in terms of the well known diagonal distributions, which have much more complicated $x$ structure, using the full Shuvaev transforms,  (\ref{eq:shuvq}) and (\ref{eq:shuvg}).

\subsection{Leading order skewed distributions}
\begin{figure}
\begin{center}
\includegraphics[width=0.49\textwidth,bb=25 5 410 360]{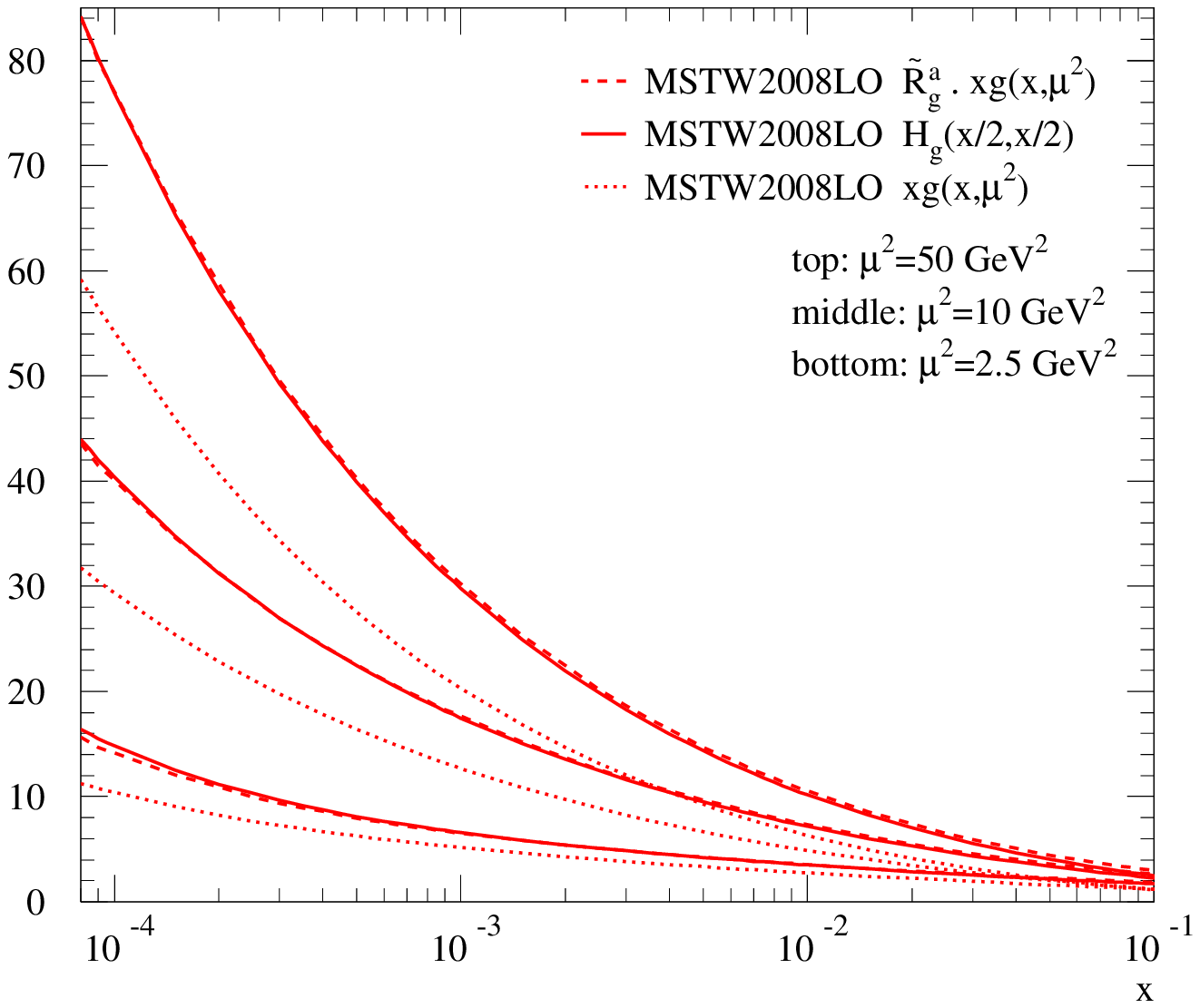}
\includegraphics[width=0.49\textwidth,bb=25 5 410 360]{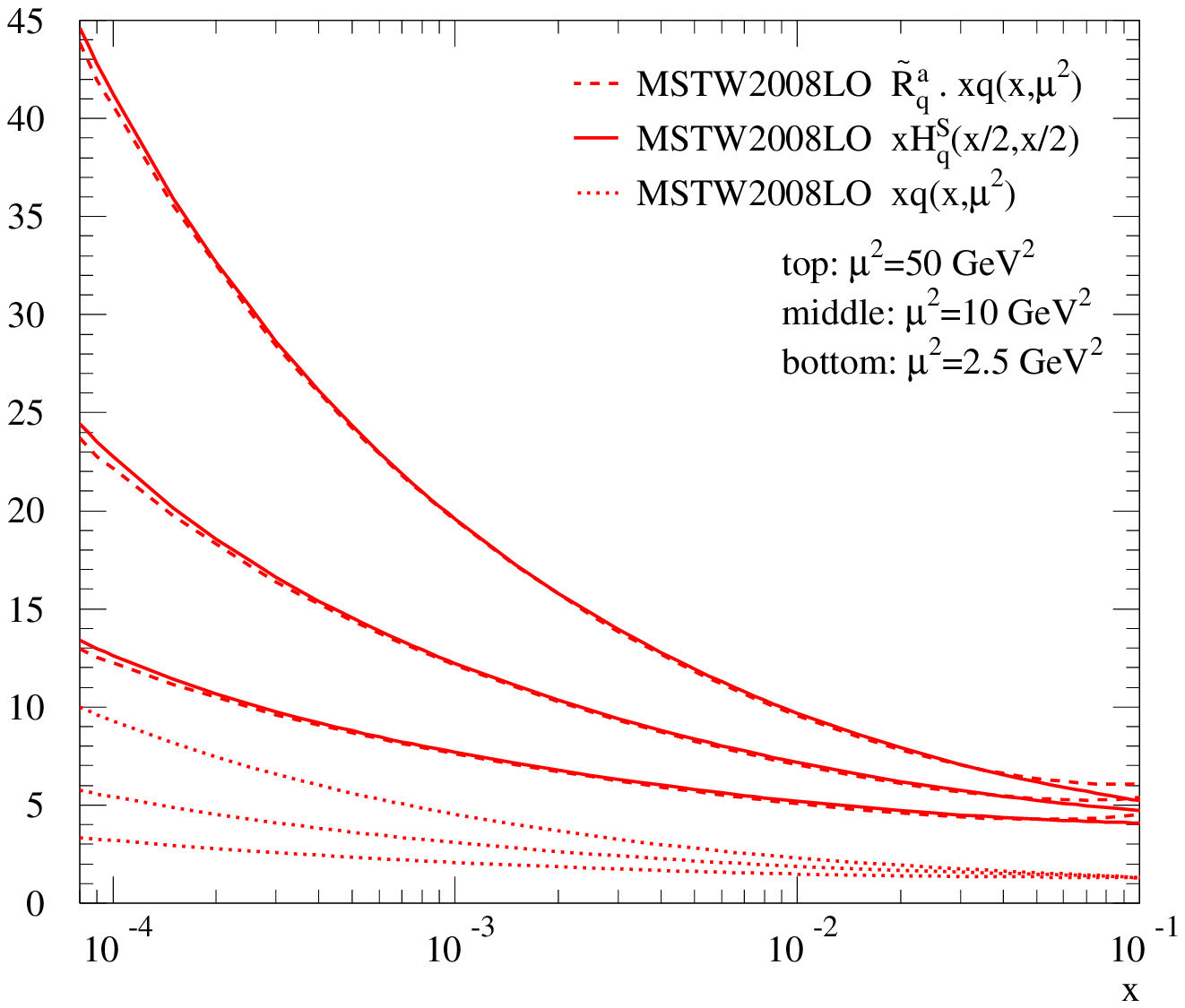}
\vspace{-5mm}
\caption{Skewed (solid and dashed) and diagonal (dotted lines) LO
  partons from MSTW2008 as a function of $x$ at scales
  $\mu^2=2.5,\,10,\,50\,\mathrm{GeV}^2$. The dashed lines 
  show the skewed partons if the analytic approximation $\tilde R^a$
  from (\protect{\ref{eq:Rtildeanalytic}}) is used. Left:
  gluons, right panel: $u, d, s$ quark singlet.} 
\label{fig:mstwlo}
\end{center}
\end{figure}
For completeness, we also calculate LO generalised
parton distributions. The LO integrated distributions have a steeper
$1/x$ behaviour than those at NLO; that is, the values of $\lambda$
are larger. Therefore the skewed effect is larger. All other features
of the LO generalised distributions are qualitatively the same as
those at NLO, see Fig.~\ref{fig:mstwlo} in comparison to
Figs.~\ref{fig:mrstandcteqgluonsx} and \ref{fig:quarksx}.

The large numerical difference between the LO and NLO integrated gluon
is due, first, to the absence of the LO coefficient function for
$\gamma^* g$ splitting, $C_{\gamma g}^{(0)}=0$; and, second, to a
singular $1/z$ term in the quark-quark splitting function $P_{qq}$
which is present at NLO, but absent at LO. To compensate for these
absences at LO, we need to `artificially' enhance the input LO gluon
at low $x$. It is, therefore, not necessarily true that the LO
contribution to another process can be effectively replaced by the
same enhanced gluons. For this reason we prefer to use NLO partons. 

Recall that the description of DVCS data in terms of the LO skewed distribution, given by the
Shuvaev approach, was not good \cite{km2}, while at NLO we observe
agreement with the data \cite{km}. The data have also been described in
terms of a dipole model \cite{Favart:2007zz}. Here, the interaction of
the gluon with the quark-antiquark pair produced by the photon plays
the role of the NLO coefficient function $C_{\gamma g}$, while the
skewed effect of the NLO gluon was calculated using the Shuvaev
prescription. Therefore, actually, the dipole approach is close to a
NLO treatment of DVCS.

\subsection{Grid interpolation package}
As the numerical computation of the full Shuvaev transform requires
some care and can be too slow for applications, we provide a simple
and fast interpolation routine in Fortran77. This includes reasonably
small (approx. 3.4 MBytes) grid files for the LO and NLO input global
analyses MSTW2008 \cite{mstw2008}, MRST2004 \cite{mrst2004} and
CTEQ6.6 \cite{cteq}. These contain
the information for quarks, antiquarks and gluons. The interpolation
routine and grid files can be downloaded from {\tt
  http://www.maths.liv.ac.uk/TheorPhys/RESEARCH/pubcodes.html}.\footnote{Grid
  files for other partons can be created upon request.} 
The grid files use 85 points in $x$ where
$4\cdot10^{-5}\le x\le 1$, together with 43 points in the ratio
$\xi/x$ where $0\le \xi/x\le 1$ and 20 points in $q^2$ where
$1.25\,\mathrm{GeV}^2\le q^2\le 80\,\mathrm{GeV}^2$. Using $\xi/x$ as
opposed to $\xi$ as a parameter increases the interpolation accuracy
as the region $\xi \lesssim x$, where the skewed distributions are
rising sharply, can be sampled more densely than regions $\xi\ll x$,
where they are flatter. The $x$ grid points were chosen on a log
scale, and the transform uses a linear interpolation on log scales for
the three parameters (apart from the bin which includes $\xi/x=0$,
where the interpolation is performed on a linear scale). The evolution in
$q^2$ is very smooth, so fewer points in $q^2$ are needed. 

The accuracy of the results from the interpolation on the grid is
always better than $0.2\%$ in the small $x$ regime (and still better than
$1\%$ for $x > 10^{-2}$), and typically much better for $\xi \ll x$ or
in the case of gluons.  Based on convergence tolerances of the double
integrals evaluated in (\ref{eq:shuvq}, \ref{eq:shuvg}), and the
numerical derivatives (of the interpolated input partons) of one per
mille, we estimate the accuracy of the grid points to be better than
$0.2\%$ in the small $x$ regime. 

Note that while in the figures above we have shown the skewed partons
only up to $x = 0.1$, the grid files contain information up to $x =
1$. Of course at large $x$ and $\xi$ we cannot justify the results,
while at small $\xi$ and large $x$ the skewed distributions are
approaching the diagonal ones. 

\section{Comparison with an alternative approach}
In general there are two possibilities to parameterise GPDs. One is to relate skewed distributions to the well known diagonal (global) parton distributions. This can be done with the help of the double distributions proposed by Radyushkin \cite{Radyushkin:1998es,dd}. The general form contains an arbitrary new function (up to the normalisation condition), but by construction it reduces to the diagonal distributions in the limit $\xi \to 0$. The Shuvaev transform is a particular case of this approach, with the advantage that after a physically reasonable assumption (that is, no singularities in the right-half $j$ plane) it gives unique GPDs in the low $x$ domain in terms of global diagonal partons, without any new parameters. The physical reasonableness of this assumption is evident from  Fig.~\ref{fig:ord} and the accompanying discussion.

An alternative approach is to fit the available data, corresponding to {\it both} skewed and diagonal distributions, using an ansatz or model for the input GPDs at some starting scale $Q_0$. This approach was used in a recent paper \cite{KM}, where the moments of the GPDs were parameterised in terms of beta ($B$) functions motivated by an SO(3) partial wave model, see Section 3.2 of \cite{KM}. In order not to lose the statistical significance of the DVCS data, which corresponds to the skewed distributions, only a small subset of global ($F_2$) data were included in the analysis. Moreover, we stress that the DVCS data do not cover the whole kinematic domain of $x$ and $\xi$, but rather correspond to the special situation $|x|=\xi$. It was emphasised in \cite{KM} that it is impossible to describe HERA data with a model with one leading SO(3) partial wave, especially at LO. Based on this observation, the authors claim that GPDs given by the Shuvaev transform are not applicable, that is, are in contradiction with the DVCS data. Note, however, that the integral corresponding to the Shuvaev transform is determined directly from global diagonal partons, which have a more complicated $x$ behaviour. It is well known that already diagonal `global' data cannot be described by a single Regge pole ansatz. Moreover global diagonal partons are poorly described at LO \cite{mstw2008}. The NLO corrections are large, since, in comparison with LO, they contain qualitatively new $1/z$ singularities in the splitting and coefficient functions. Furthermore, note that the analytic formula for the ratio\footnote{It is misleading to quote the ratio  as ``conformal'' (as was done for eq.(28) of \cite{KM}) since it relies on a single Regge pole ansatz, as well as on conformal symmetry.} (\ref{eq:ranalytic}), which indeed is written for one Regge pole, was given just to qualitatively illustrate the discussion.

Going to NLO, we see no contradiction of the integral, parameter-free, Shuvaev transform approach with the available low $x$ DVCS data \cite{Favart:2007zz}. Moreover, the ratios $r$ of skewed-to-diagonal PDFs given in Fig.~7 of \cite{KM} for the NLO $\overline{\rm MS}$ scheme are very close to those that we obtain using the Shuvaev transform with recent global partons. Table \ref{tab:comp} shows the agreement between the NLO $\overline{\rm MS}$ GPDs of \cite{KM} and our values obtained using the Shuvaev transform at $x=0.001$ for $Q^2=10$ and $50~ {\rm GeV}^2$. It is clear from the Table that the difference between the GPDs based on the Shuvaev transform and those obtained in \cite{KM} by the NLO $\overline{\rm MS}$ fit to the `skewed' DVCS data, is much less than the DVCS error bars.

\begin{table}[h]
\begin{center}
\begin{tabular}{|c|c|c|c|c|}
\hline
& \multicolumn{2}{|c|}{$Q^2 = 10\,\GeV^2$} & \multicolumn{2}{|c|}{$Q^2 = 50\,\GeV^2$}\\
\cline{2-5}
& $~~r_G~~$ & $r_{QS}$ & $~~r_G~~$ & $r_{QS}$\\
\hline
CTEQ6.6M & 1.04 & 1.67 & 1.06 & 1.71 \\
MSTW2008NLO & 1.03 & 1.65 & 1.05 & 1.70 \\
NLO $\overline{\rm MS}$ GPDs of \cite{KM} & 1.06 & 1.76 & 1.05 & 1.77 \\
\hline
\end{tabular}
\caption{Comparison of the skewed-to-diagonal ratio for gluons
  $r_G=H_g(x,x)\,/\,H_g(x,0)$ and $u,d,s$ quark singlet
  $r_{QS}=H_q^S(x,x)\,/\,xH_q^S(x,0)$ at $x=0.001$ for scales $Q^2=10$
  and $50~ {\rm GeV}^2$.  Diagonal CTEQ6.6M and MSTW2008NLO partons
  are used as input to calculate the skewed distributions, and the
  resulting ratios are compared to the ratios as displayed in Fig.~7
  of \cite{KM} using their NLO $\overline{\rm MS}$ ($\sum$-PW) GPDs.}
\label{tab:comp}
\end{center}
\end{table}

\section{Conclusions}
We have shown how the skewed (or generalised) parton distributions,
$H_i(x,\xi)$, that are needed to describe diffractive processes, can be
obtained, for small $x$, {\it directly} from the conventional integrated global parton
densities, $f_i(x)$. This parameter-free method is based on the Shuvaev transform, and
is applicable to accuracy\footnote{The accuracy is $O(\xi^2)$ at LO.}
$O(\xi)$ at NLO, which should be sufficient for the description of all
diffractive processes of interest. For the reasons given in the
previous section the alternative approach of Ref.~\cite{KM} does not
have the advantage of being parameter free.

First we identified the reason for the doubts which had been raised
concerning the use of the Shuvaev transform. Further, we noted that
the Shuvaev transform follows from the physically-motivated conjecture
that the small $x$ input is specified by Regge physics, and so does
not generate any singularities in the right-half ($j>1$) plane in the
space-like ($x>\xi$) domain. The transform therefore has
practical applicability. 
We then investigated the
kinematic range of reliability of the simplified analytic formula
(\ref{eq:ranalytic}). This formula was derived for $x=\xi$, assuming
that the unintegrated parton distributions $xf_i$ have a power-like
$x^{-\lambda_i}$ behaviour. The `analytic' formula was only introduced to illustrate some of the qualitative features of GPDs. 

In order to obtain {\it accurate}
generalised parton distributions at NLO at {\it any} small $x, \xi$, we performed a
detailed computation of the integrals of the Shuvaev
transform. Finally, these computations allow us to provide a readily
accessible package which allows the evaluation of GPDs for arbitrary
small $x, \xi$ in the space-like $(x>\xi)$ domain. These NLO GPDs should facilitate the theoretical description of {\it all} diffractive processes of interest.

\section*{Appendix: The inverse transform}
Although we do not need the inverse transformation for the calculation
of the skewed distributions for small $\xi$ from diagonal partons
using the Shuvaev transform, it is still worth noting that the inverse
problem is solvable. Here we shall briefly clarify the inverse
transformation and show that the regions $|x|>|\xi|$ and $|x|<|\xi|$
transform separately.

Our aim is to find the function $f(x)$ whose Mellin moments,
$$
M_N\,=\,\int \frac{{\rm d}x}{x}\,x^N f(x),
$$
are equal to Gegenbauer moments of the GPD $H(x,\xi)$,
$$
G_N(\xi)\,=\,\frac{N!\Gamma(3/2)}{2^N\Gamma(N+3/2)}
\int {\rm d}x\,\xi^N C_N^{\frac 32}(x/\xi)\,H(x,\xi),
$$
$G_N(\xi)\,=\,M_N$. Here for definiteness we consider the non-singlet channel,
the singlet one can be treated in the same manner.

With the generating function for Gegenbauer
polynomials,
$$
\xi^N C_N^{\frac 32}(x/\xi)\,=\,\frac 1{N!}\frac{\partial^N}{\partial t^N}
R^{-\frac 32}(t,x)\biggl|_{t=0},
$$
$$
R(t,x)\,=\,1\,-\,2\,tx\,+\,t^2\xi^2,
$$
the moments take the form
$$
G_N(\xi)\,=\,\frac 12\,\int_0^1 {\rm d}y \int_0^1 {\rm d}v(1-v)^{-\frac 12}
\frac 1{2\pi i}\oint\frac{dt}{t^{N+1}}
R^{-\frac 32}\bigl(\frac 12vt,y\bigr)\,H(y,\xi),
$$
where the contour of the $t$-integration goes in the complex plane
around $t=0$, while the integral over $v$ yields the normalisation.
Integrating by parts over $y$ (this is needed to make the kernel
less singular) we rewrite this expression as
\begin{eqnarray}
G_N(\xi)\,&=&\,-\int_0^1 {\rm d}y\int_0^1\frac{dv}{v}(1-v)^{-\frac 12}
\frac 1{2\pi i}\oint\frac{dt}{t^{N+1}}\frac 1t
R^{-\frac 12}\bigl(\frac 12vt,y\bigr)\,\frac{\partial}{\partial y}H(y,\xi)
\nonumber \\
&&+\,\int_0^1\frac{{\rm d}v}{v}(1-v)^{-\frac 12}
\frac 1{2\pi i}\oint\frac{dt}{t^{N+1}}\frac 1t
R^{-\frac 12}\bigl(\frac 12vt,y\bigr)\,H(y,\xi)\biggr|_{-1}^1.
\nonumber
\end{eqnarray}
The contour of $t$-integration can be shrunk to
the cut of the radical in the integrand,
\begin{eqnarray}
-\frac 1{2\pi i}\oint\frac{dt}{t^{N+1}}\frac 1t
R^{-\frac 12}\bigl(\frac 12vt,y\bigr)\,&=&\,
\frac 1\pi \int_{t_2}^{t_1}\frac{{\rm d}t}{t^{N+1}}\frac 1t
R_c^{-\frac 12}(t,y) \nonumber \\
\,&=&\,-\,\frac 1\pi \int_{\frac 1{t_2}}^{\frac 1{t_1}}\frac{{\rm d}x}{x}\,x^{N+1}
R_c^{-\frac 12}(\frac 1x,y),\nonumber
\end{eqnarray}
where the values $t_{1,2}=2/(\xi^2v)[\,y\pm \sqrt{y^2-\xi^2}\,]$ determine
the real (for $|y|>|\xi|$) interval, in which
$$
R_c(t,y)\,=\,vyt-\frac 14 \xi^2v^2t^2-1>0.
$$
These relations show that the desired function is given by the integral
transformation
$$
f(x)\,=\,\int_0^1 {\rm d}y\,K(x,y)\bigl[\frac{\partial}{\partial y}\,+\,
\delta(y-1)-\delta(y+1)\bigr]H(y,\xi),
$$
with the kernel
$$
K(x,y)\,=\,-\frac 1\pi\int_0^1\frac{{\rm d}v}{v}(1-v)^{-\frac 12}
\bigl[x^2\,Y^{-\frac 12}\theta(Y)\bigr],
$$
$$
Y\,\equiv\,vxy-\frac 14\xi^2v^2-x^2,\qquad |y|>|\xi|.
$$
Note that for $|\xi|\ll 1$ the transformation becomes close to the identity,
$f(x)=H(x)+{\cal O}(\xi)$.

If $|y|<|\xi|$ the branch points $t_{1,2}$ are complex. In this case
the function $f$ will be defined on the segment lying in the complex plane.
Thus the inverse transform does not exist as a real function for the
whole interval.

\section*{Acknowledgements}
We thank Markus Diehl, Dieter M{\"u}ller and Graeme Watt for valuable
discussions, and Graeme Watt for the provision of the MSTW2008 partons
prior to publication.

\end{document}